\documentclass[sigconf,pbalance]{acmart}

\AtBeginDocument{%
  }

\usepackage{pifont}
\usepackage{multirow}
\usepackage{url}

\copyrightyear{2026}
\acmYear{2026}
\setcopyright{cc}
\setcctype{by}
\acmConference[ICPC '26]{34th IEEE/ACM International Conference on Program Comprehension}{April 12--13, 2026}{Rio de Janeiro, Brazil}
\acmBooktitle{34th IEEE/ACM International Conference on Program Comprehension (ICPC '26), April 12--13, 2026, Rio de Janeiro, Brazil}
\acmPrice{}
\acmDOI{10.1145/3794763.3794791}
\acmISBN{979-8-4007-2482-4/2026/04}

\begin{document}

\title[Test Behaviors, Not Methods! Detecting Tests Obsessed by Methods]{Test Behaviors, Not Methods!\\Detecting Tests Obsessed by Methods}

\author{Andre Hora}
\orcid{0000-0003-4900-1330}
\affiliation{%
  \institution{Department of Computer Science, UFMG}
  \city{Belo Horizonte}
  \country{Brazil}
}
\email{andrehora@dcc.ufmg.br}

\author{Andy Zaidman}
\orcid{0000-0003-2413-3935}
\affiliation{%
  \institution{Delft University of Technology}
  \city{Delft}
  \country{The Netherlands}
}
\email{a.e.zaidman@tudelft.nl}

\begin{abstract}
Best testing practices state that tests should verify a single functionality or behavior of the system.
Tests that verify multiple behaviors are harder to understand, lack focus, and are more coupled to the production code.
An attempt to identify this issue is the test smell \emph{Eager Test}, which aims to capture tests that verify too much functionality based on the number of production method calls.
Unfortunately, prior research suggests that counting production method calls is an inaccurate measure, as these calls do not reliably serve as a proxy for functionality.
We envision a complementary solution based on runtime analysis: we hypothesize that some tests that verify multiple behaviors will likely cover multiple paths of the same production methods.
Thus, we propose a novel test smell named \emph{Test Obsessed by Method}, a test method that covers multiple paths of a single production method.
We provide an initial empirical study to explore the presence of this smell in 2,054 tests provided by 12 test suites of the Python Standard Library.
(1) We detect 44 \emph{Tests Obsessed by Methods} in 11 of the 12 test suites.
(2) Each smelly test verifies a median of two behaviors of the production method.
(3) The 44 smelly tests could be split into 118 novel tests.
(4) 23\% of the smelly tests have code comments recognizing that distinct behaviors are being tested.
We conclude by discussing benefits, limitations, and further research.
\end{abstract}

%
%




\maketitle

\section{Introduction}

Best testing practices state that test methods should verify a single functionality or behavior of the system~\cite{winters2020software, meszaros2007xunit, Moonen2008, van2001refactoring, anicheTSE2022}.
The Google Testing Blog refers to this practice as \emph{``test behaviors, not methods''}~\cite{behaviors}.
This simple, but powerful recommendation brings several benefits to software development.
First, tests are focused and easier to understand since each test contains code to exercise only one behavior~\cite{winters2020software}.
Second, it reduces the coupling between the test and production code~\cite{winters2020software}.
Third, when a new behavior is added to the system, a new test should be created for \emph{that} behavior (i.e.,~existing tests are not changed).
Thus, tests are more resilient to changes since adding new behaviors is unlikely to break the existing tests~\cite{greilerMSR2013, winters2020software}.

\begin{figure}[t]
     \centering
         \includegraphics[width=0.4\textwidth]{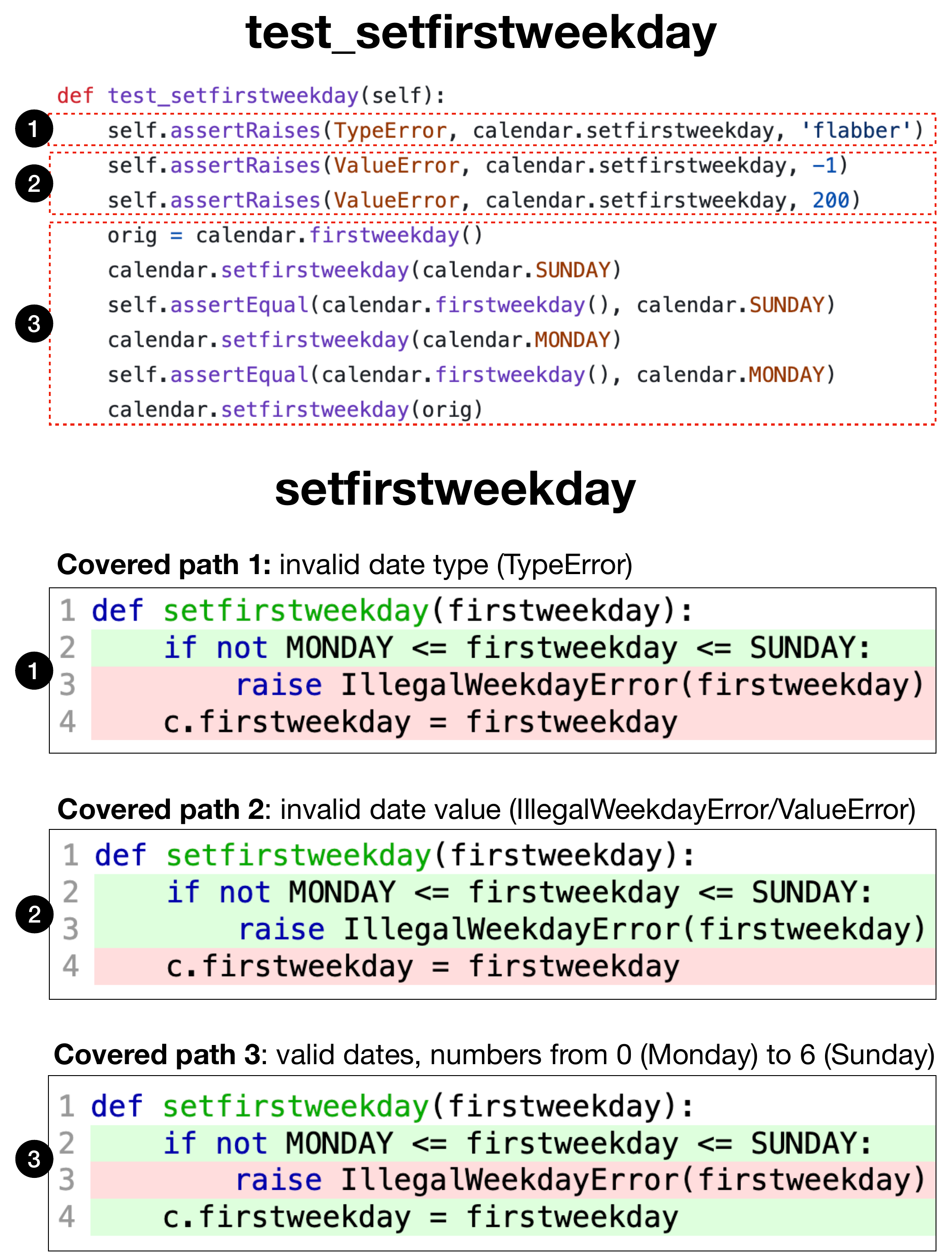}  
         \vspace{-3mm}
         \caption{Example of a \emph{Test Obsessed by Method} in CPython. Test \texttt{test\_setfirstweekday} covers three paths of \texttt{setfirstweekday}. This test could be split into three tests.}
         \Description{Example of a Test Obsessed by Method in CPython.}
         \label{fig:overview}
         \vspace{-4mm}
\end{figure}

Identifying test methods that violate this best practice is important for uncovering tests that verify multiple behaviors, that is, tests that are harder to understand, lack focus, are more coupled to the production code, and are less resilient to changes.
However, this is not a trivial task because functionality (or behavior) is hard to define.
An attempt at solving this is the test smell \emph{Eager Test}, which aims to capture tests that verify too much functionality~\cite{meszaros2007xunit, van2001refactoring}.
Most solutions and tools to catch this test smell rely on rules that count the number of production method calls in test methods as a proxy for ``the number of functionalities''~\cite{van2007detection, grano2019scented, peruma2020tsdetect, panichella2022test}.
Unfortunately, it is not ideal to count production method calls to detect tests that verify multiple functionalities.
Prior studies report it can be inaccurate~\cite{panichella2022test}, has limited predictive power~\cite{van2007detection}, and developers tend to disagree with low threshold values of method calls~\cite{spadini2020investigating}.

Given the aforementioned limitations of properly detecting test methods that verify multiple behaviors, we envision a complementary solution based on runtime analysis.
\emph{We hypothesize that some test methods that verify multiple behaviors will likely cover multiple paths of the same production methods}.
Thus, we propose to use the covered paths of the production methods as a proxy for behaviors (see Figure~\ref{fig:overview} for an example).
Unlike \emph{Eager Test} implementations that work with static analysis (i.e.,~count of method calls), this novel smell works with runtime analysis (i.e.,~count of covered paths).

Thus, we propose a novel test smell named \emph{Test Obsessed by Method}, a test method that covers multiple paths of a single production method.
A test smell can be seen as a symptom of a problem~\cite{meszaros2007xunit}, and in this case, it manifests as the test method's greediness in trying to cover multiple paths of a production method, leading to a test that potentially verifies multiple behaviors.
Figure~\ref{fig:overview} presents an example of a \emph{Test Obsessed by Method} in CPython.
The test \texttt{test\_\-set\-first\-weekday}\footnote{test\_\-set\-first\-weekday: \url{https://github.com/python/cpython/blob/2938c3d/Lib/test/test_calendar.py\#L513}} 
calls two production methods (\texttt{firstweekday} and \texttt{set\-first\-weekday}), but the issue is that it
verifies multiple behaviors of \texttt{set\-first\-weekday}.
Specifically, the test executes three distinct paths of \texttt{set\-first\-weekday} (as detailed in Figure~\ref{fig:overview}).
Path 1 is executed due to an invalid date type, which raises the exception \texttt{Type\-Error}.
Path 2 occurs due to the invalid date values, resulting in the exception \texttt{Illegal\-Weekday\-Error}.
Path 3 is executed due to the valid dates (i.e.,~0 to 6).
This test could be fixed by splitting it into three test methods, one for each behavior, as recommended by best testing practices~\cite{winters2020software}.

This paper has two contributions.
First, we propose a novel test smell named \emph{Test Obsessed by Method} (Section~\ref{sec:test_smell}).
Second, we provide an initial empirical study to explore the presence of \emph{Tests Obsessed by Methods} in real-world test suites (Section~\ref{sec:study}).
We analyze 2,054 test methods of 12 real-world test suites of the Python Standard Library.
To detect the smells, we run an instrumented version of the test suites and collect information about the executed lines of code at runtime.
We define two research questions:

\begin{itemize}
    \item \textbf{RQ1: How prevalent are Tests Obsessed by Methods}?
    We detect 44 \emph{Tests Obsessed by Methods} in 11 of 12 test suites.
    Each smelly test verifies two behaviors of the production method on the median.
    
    \item \textbf{RQ2: How can we fix Tests Obsessed by Methods}?
    The 44 smelly tests could be split into 118 novel tests.
    We find code comments in 10 out of 44 (23\%) smelly tests, recognizing that distinct behaviors are being tested.
    
\end{itemize}

Finally, we conclude the paper by discussing the benefits and limitations of the proposed test smell and further research.

\section{Related Work}

Ideally, test suites should have good quality to catch bugs and protect against regressions~\cite{anicheTSE2022, hora2024test, hora2025exceptional, dalton2020exceptional, marcilio2021java}.
Test smells indicate potential design problems in the test code~\cite{van2001refactoring, meszaros2007xunit}.
The presence of test smells in test suites may affect the test quality, maintainability, and extendability, reducing their effectiveness in finding bugs in production code~\cite{bavota2012empirical, peruma2020tsdetect, spadini2018relation, bavota2015test, garousi2018smells, tufano2016empirical, athanasiouTSE2014}.
Ideally, test methods should verify a single functionality or behavior of the system~\cite {winters2020software}.
Tests that violate this best practice are considered \emph{eager}.
According to Meszaros~\cite{meszaros2007xunit}, an \emph{Eager Test} is a test that verifies too much functionality.
Van Deursen~\emph{et al.}~\cite{van2001refactoring, Moonen2008} originally defined it as a test that checks several methods of the object to be tested.
Both definitions~\cite{meszaros2007xunit, van2001refactoring} address the \emph{Eager Test} informally; the number of verified functionalities or methods is not clear.

To overcome this limitation, other studies define \emph{Eager Test} more formally~\cite{van2007detection, grano2019scented, peruma2020tsdetect, panichella2022test}.
For example, Van Rompaey~\emph{et al.}~\cite{van2007detection} proposed a metric-based approach that relies on the number of production method calls, but concluded that this metric has limited predictive power.
Test smell detection tools also rely on the production method calls to detect \emph{Eager Tests}, for example, setting a threshold of at least \emph{two} calls to production methods~\cite {grano2019scented, peruma2020tsdetect}.
Spadini~\emph{et al.}~\cite{spadini2020investigating} reported that developers disagree with such a low threshold, finding that \emph{four} calls to production methods are better suited to detect \emph{Eager Tests}.
Recently, Panichella~\emph{et al.}~\cite{panichella2022test} analyzed two test smell detection tools~\cite{grano2019scented, peruma2020tsdetect} and concluded that existing tools simply rely on rules that count the number of production method calls in test methods as a proxy for ``the number of functionalities'', suggesting that such a simple heuristic is highly inaccurate.
Another important conclusion of the research is that it is non-trivial to detect \emph{Eager Tests} automatically, and it is fault-prone to assume a threshold of just two calls.
Thus, detecting test methods that verify multiple functionalities requires more semantic awareness than is currently considered~\cite{panichella2022test}.


\section{Tests Obsessed by Methods}
\label{sec:test_smell}

\subsection{Overview}

A \emph{Test Obsessed by Method} is a test method that covers multiple paths of a single production method.
The rationale is that each covered path in such a production method represents a behavior.
Thus, a \emph{Test Obsessed by Method} is a test that potentially verifies multiple behaviors.
To fix a test method with this smell, we can create one test method for each tested behavior.
In this case, behaviors can be conveniently identified by the covered paths of the production method.
For example, the production method \texttt{set\-first\-weekday} presented in Figure~\ref{fig:overview} has three covered paths, thus, the test \texttt{test\_set\-first\-weekday} could be split into three tests.

\begin{figure}[h]
     \centering
         \vspace{-3mm}
         \includegraphics[width=0.4\textwidth]{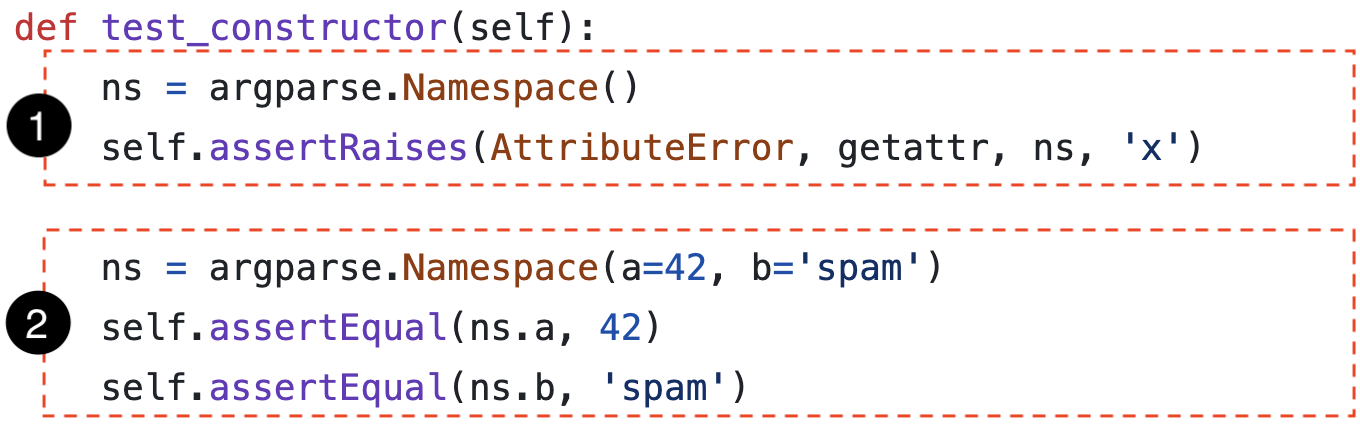}
         \vspace{-3mm}
         \caption{\emph{Test Obsessed by Method}: \texttt{test\_constructor} tests two behaviors of \texttt{Namespace} (argparse).}
         \Description{\emph{Test Obsessed by Method}: \texttt{test\_constructor} tests two behaviors of \texttt{Namespace} (argparse).}
        \label{fig:example1}
\end{figure}

Figure~\ref{fig:example1} presents \texttt{test\_\-constructor}\footnote{test\_constructor: \url{https://github.com/python/cpython/blob/f474391b/Lib/test/test_argparse.py\#L5673}} of the CPython argparse library. 
This test is problematic because it verifies two behaviors of the production method \texttt{Namespace}.
These behaviors should ideally be tested in two distinct test methods to verify: (1) an exceptional case that raises the exception \texttt{AttributeError} and (2) valid cases.

\begin{figure}[h]
     \centering
         \includegraphics[width=0.45\textwidth]{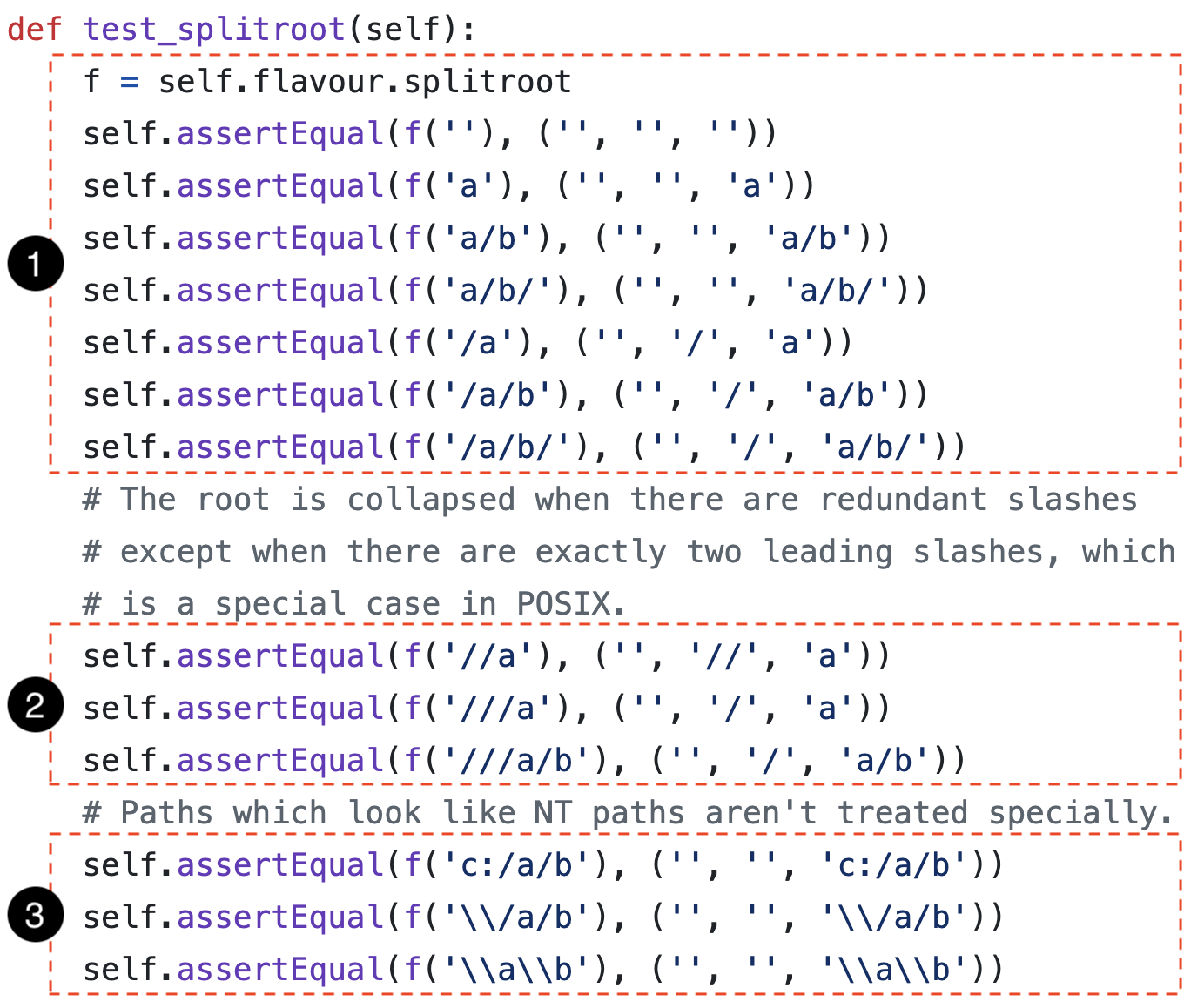}
         \vspace{-3.5mm}
         \caption{\emph{Test Obsessed by Method}: \texttt{test\_splitroot} tests three behaviors of \texttt{splitroot} (pathlib).}
         \Description{\emph{Test Obsessed by Method}: \texttt{test\_splitroot} tests three behaviors of \texttt{splitroot} (pathlib).}
        \label{fig:example2}
\end{figure}

Figure~\ref{fig:example2} shows the test \texttt{test\_\-split\-root}\footnote{test\_splitroot: \url{https://github.com/python/cpython/blob/0c5fc272/Lib/test/test_pathlib.py\#L80}} of the CPython pathlib library. 
This test method is problematic because it verifies three behaviors of the production method \texttt{split\-root}.
These behaviors should ideally be tested in three distinct test methods, covering (1) basic paths, (2) POSIX paths (i.e.,~Unix-like), and (3) NT paths (i.e.,~Windows).
Interestingly, the code comments on the test method highlight that distinct requirements are being tested.

\begin{figure}[h]
     \centering
         \includegraphics[width=0.47\textwidth]{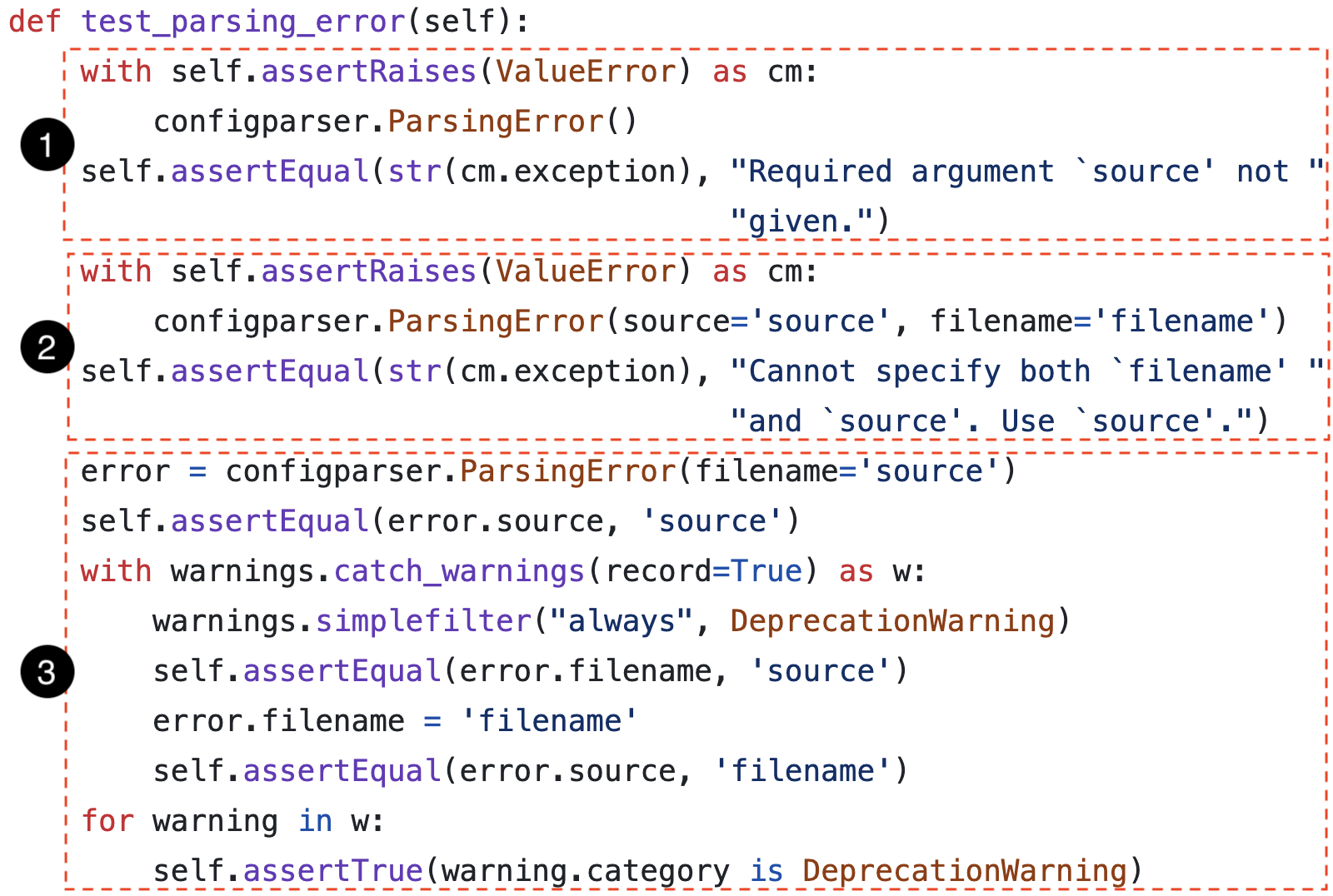}
         \vspace{-3.5mm}
         \caption{\emph{Test Obsessed by Method}: \texttt{test\_parsing\_error} tests three behaviors of \texttt{ParsingError} (configparser).}
         \Description{\emph{Test Obsessed by Method}: \texttt{test\_parsing\_error} tests three behaviors of \texttt{ParsingError} (configparser).}
        \label{fig:example3}
\end{figure}

Lastly, Figure~\ref{fig:example3} presents \texttt{test\_\-split\-root}\footnote{test\_parsing\_error: \url{https://github.com/python/cpython/blob/0c5fc272/Lib/test/test_configparser.py\#L1604}} of the CPython configparser library. 
This test is problematic because it verifies multiple exceptional behaviors~\cite{hora2025exceptional, dalton2020exceptional, marcilio2021java} of the production method \texttt{Parsing\-Error}.
These behaviors should ideally be tested in three distinct test methods, covering the distinct forms to use \texttt{Parsing\-Error}.

It is important to note that a test method calling the same production method multiple times does not necessarily pose a problem.
Consider, for example, a test method that calls method \texttt{add} of some data structure multiple times and then verifies the data structure size.
In this case, calling the same method \texttt{add} multiple times is not an issue, and this test method is not a \emph{Test Obsessed by Method}.
Indeed, test methods like this one will be present in any test suite; thus, the challenge is to distinguish such valid tests from the ones that are \emph{really} verifying multiple behaviors of a production method.
To reduce the possibility of false positives, we rely on runtime analysis to detect \emph{Tests Obsessed by Methods}.

\subsection{Detecting Tests Obsessed by Methods}

To detect \emph{Tests Obsessed by Methods}, we perform runtime analysis by collecting data during test execution.
We collect the covered paths of every production method executed by the test methods.
A covered path refers to a set of input values that cause the production method to follow the same execution flow, resulting in the execution of identical lines of code.
If a test covers two or more paths of a production method, it is classified as smelly. 
For example, \texttt{test\_\-setfirst\-weekday} covers three paths of \texttt{set\-first\-weekday}: (1) line 2; (2) lines 2,3; and (3) lines 2,4; thus, it is smelly.

\section{Preliminary Empirical Study}
\label{sec:study}

\subsection{Design}

\noindent\textbf{Case Study}:
We aim to identify the presence of \emph{Tests Obsessed by Methods} in real-world test suites.
For this purpose, we analyze 2,054 test methods of 12 test suites of the Python Standard Library: gzip, email, calendar, ftplib, collections, os, tarfile, pathlib, logging, smtplib, argparse, and configparser.
Our dataset is available at: \url{https://doi.org/10.5281/zenodo.17469070}.

\noindent\textbf{Runtime Analysis}:
To detect the covered paths of the production methods, we run an instrumented version of the test suites and collect information about the executed lines at runtime.
We rely on SpotFlow~\cite{spotflow}, a tool to ease runtime analysis in Python, which is implemented with the support of the standard trace function~\cite{python_sys_trace}.



\subsection{Results}

\noindent\textbf{RQ1: How prevalent are Tests Obsessed by Methods?}
We find 54 \emph{Test Obsessed by Method} candidates in the 2,054 analyzed test methods.
We manually analyzed the 54 tests and detected 44 true positives and 10 false positives, resulting in a precision of 81.5\%.
True positives are test methods that can be split into multiple tests, while false positives are harder or impossible to split. 
Examples of true positives are presented in Figures~\ref{fig:overview}-\ref{fig:example3}.
False negatives occur in scenarios where the production method is invoked indirectly within the test or is executed as part of the test setup.
The smelly tests are present in 11 of 12 test suites; on the median, they exercise two paths of the production method.
Table~\ref{tab:rq1} details the number of \emph{Tests Obsessed by Methods} by covered paths.


\begin{table}[h]
\centering
\small
\caption{\emph{Tests Obsessed by Methods} by covered paths.}
\vspace{-3.5mm}
\begin{smaller}
\begin{tabular}{l c ccccc}
\toprule
& \multirow{2}{*}{\textbf{Total}} & \multicolumn{5}{c}{\textbf{Covered Paths}} \\ \cline{3-7}
& & 2 & 3 & 4 & 5 & 7 \\ \midrule
\textbf{\#Tests Obsessed by Methods} & 44 & 25 & 12 & 5 & 1 & 1 \\
\bottomrule
\end{tabular}
\end{smaller}
\label{tab:rq1}
\end{table}


\noindent\textbf{RQ2: How can we fix Tests Obsessed by Methods?}
In this RQ, we explore how the \emph{Tests Obsessed by Methods} could be potentially fixed.
Considering that each smelly test verifying \emph{n} behaviors could be split into \emph{n} tests (i.e.,~one for each covered path), the 44 smelly tests could be split into 118 novel tests (i.e.,~25$\times$2 $+$ 12$\times$3 $+$ 5$\times$4 $+$ 1$\times$5 $+$ 1$\times$7), as detailed in Table~\ref{tab:rq1}.

Interestingly, among the 44 smelly tests, 10 have code comments recognizing that distinct behaviors are, in fact, being tested.
Due to the space limit, we briefly present three examples.
The first example happens in the test \texttt{test\_\-split\-root} (see Figure~\ref{fig:example2}).
In this case, the comments highlight that distinct behaviors of the production method \texttt{splitroot} are tested: (1) basic paths, (2) POSIX paths, and (3) NT paths.
The second example is the test \texttt{test\_\-is\_\-tarfile\_\-erroneous}\footnote{test\_is\_\-\-tarfile\_\-erroneous: \url{https://github.com/python/cpython/blob/850189a6/Lib/test/test_tarfile.py\#L359}} of the tarfile library, which tests two behaviors of \texttt{is\_tarfile}.
In this case, the comments suggest that two behaviors of \texttt{is\_tarfile} are verified: (1) for string tar files and (2) for file-like object tar files.
Finally, the third example happens in the test \texttt{test\_\-is\_\-absolute}\footnote{test\_is\_absolute: \url{https://github.com/python/cpython/blob/850189a6/Lib/test/test_pathlib.py\#L1222}} of the pathlib library.
It tests two behaviors of the method \texttt{is\_\-absolute}: (1) for NT files and (2) for UNC paths.

\begin{center}
\fcolorbox{black}{gray!15}{
  \parbox{\dimexpr1\linewidth-2\fboxsep-2\fboxrule}{
\textbf{Summary}:
(1) We detect 44 \emph{Tests Obsessed by Methods} in 11 of 12 test suites.
(2) Each smelly test verifies a median of two behaviors of the production method.
(3) The 44 smelly tests could be split into 118 novel tests.
(4) 10 in 44 (23\%) smelly tests have comments recognizing that distinct behaviors are being tested.
  }
}
\end{center}


\section{Discussion}
\label{sec:discussion}

\subsection{Fixing Tests Obsessed by Methods}
We identified \emph{Tests Obsessed by Methods} in 11 of the 12 analyzed test suites, indicating that the problem is spread over multiple projects rather than isolated.
One important aspect of \emph{Tests Obsessed by Methods} is that behaviors can be conveniently identified by the covered paths of the production method.
This is why the 44 smelly tests could be refactored into 118 novel, focused tests.
In contrast, the 10 false positives refer to cases harder to refactor, such as production methods part of the test setup.

Thus, the proposed test smell not only identifies the problem (i.e.,~testing multiple behaviors within a single test), but also provides guidance for resolution (i.e.,~create one test per covered path).

\subsection{Runtime Analysis}

Most test smell detection techniques rely on static analysis.
However, identifying \emph{Tests Obsessed by Methods} requires runtime (dynamic) analysis, which involves executing the test suite and gathering runtime data.
Other test smells also rely on runtime analysis, for example, \emph{Rotten Green Tests}, which are passing tests with at least one assertion not executed~\cite{delplanque-2019-rotten, aranega2021rotten, martinez2020rtj, robinson2023rotten}.
While runtime analysis is more expensive than static, it can identify problems that are only ``visible'' when code is executed, such as a test with an assertion not executed or that covers multiple paths of a production method.

\subsection{Comparison with Eager Test}

\emph{Eager Test} is defined as a test method that contains multiple calls to multiple production methods~\cite{peruma2020tsdetect}.
It is unclear what should be the minimum number of production method calls: some studies suggest at least two calls~\cite{grano2019scented, peruma2020tsdetect}, while others recommend at least four calls~\cite{spadini2020investigating}.
Moreover, constructor calls are typically excluded~\cite{panichella2022test}; otherwise, any test that instantiates a class and calls a method could be an \emph{Eager Test}.
Considering the six \emph{Tests Obsessed by Methods} discussed in this paper, only \texttt{test\_setfirstweekday} with a threshold of 2 calls would be classified as \emph{Eager Test} as well.
All the other tests would not be classified as \emph{Eager Test}, as detailed in Table~\ref{tab:comparison}.
Therefore, \emph{Tests Obsessed by Methods} can detect smelly tests that \emph{Eager Test} does not identify.

\begin{table}[!t]
\centering
\small
\caption{\emph{Eager Test} vs. \emph{Test Obsessed by Method}.}
\vspace{-3.5mm}
\begin{smaller}
\begin{tabular}{l ccc}
\toprule
\multirow{2}{*}{\textbf{Test Method}} & \multicolumn{2}{c}{\textbf{Eager Test}} & \textbf{Test Obsessed} \\ \cline{2-3}
& 2 calls & 4 calls & \textbf{by Method} \\ \midrule

\texttt{test\_setfirstweekday} & \ding{52} & \ding{56} & \ding{52} \\
\texttt{test\_constructor} & \ding{56} & \ding{56} & \ding{52} \\
\texttt{test\_splitroot} & \ding{56} & \ding{56} & \ding{52} \\
\texttt{test\_parsing\_error} & \ding{56} & \ding{56} & \ding{52} \\
\texttt{test\_is\_tarfile\_erroneous} & \ding{56} & \ding{56} & \ding{52} \\
\texttt{test\_is\_absolute} & \ding{56} & \ding{56} & \ding{52} \\

\bottomrule
\end{tabular}
\end{smaller}
\label{tab:comparison}
\end{table}

\section{Limitations}

The proposed test smell is not intended to detect \emph{all} test methods that verify multiple behaviors.
Instead, we aim to identify \emph{some} test methods that verify multiple behaviors, particularly those focused on testing multiple behaviors of a single production method.
Therefore, it should be used in complement to other test smells, such as \emph{Eager Test}, rather than a substitute.
In fact, \emph{Tests Obsessed by Methods} may identify smelly tests that \emph{Eager Test} misses, and vice versa.
Further analysis is needed to better compare both smells.

\section{Conclusion and Future Work}

We proposed a novel test smell named \emph{Test Obsessed by Method}, a test method that covers multiple paths of a single production method.
We conducted an initial study to explore the presence of this smell and found it in 11 of 12 test suites.

\smallskip

\noindent\textbf{Future Work:}
First, we plan to conduct a qualitative study with experts to better understand the limitations of tests that verify multiple behaviors.
Second, we plan to expand the empirical study by including more real-world test suites, not only from the Python Standard Library.
Lastly, we intend to conduct a contribution study~\cite{brandt2024shaken, hora2024pathspotter, danglot2019automatic} in which we submit pull requests containing refactorings that remove \emph{Tests Obsessed by Methods} in open-source projects.
We hope to better understand software engineers' ideas about the problem and the proposed refactoring.



\begin{acks}
This research was supported by CNPq (process 403304/2025-3), CAPES, and FAPEMIG.
This work was partially supported by INES.IA (National Institute of Science and Technology for Software Engineering Based on and for Artificial Intelligence), www.ines.org.br, CNPq grant 408817/2024-0.
\end{acks}

\bibliographystyle{ACM-Reference-Format}
\bibliography{main}

\end{document}